\documentclass[runningheads]{llncs}

\usepackage{graphicx}
\usepackage{comment}
\usepackage{booktabs}
\usepackage{multirow}
\usepackage{array}
\usepackage{caption}
\usepackage{subfiles}
\usepackage{subfig}
\usepackage{ragged2e}
\usepackage{todonotes}
\usepackage{hyperref}
\usepackage{enumitem}
 \usepackage{makecell}
\usepackage{tabularx}
\usepackage{adjustbox}
\usepackage{comment}
\usepackage{graphicx}
\usepackage{pifont}
\usepackage{tablefootnote}
\usepackage{threeparttable}
\usepackage{cite}
\usepackage[misc,geometry]{ifsym}
\usepackage{hyperref} 

\newcolumntype{L}{>{\centering\arraybackslash}m{2.3cm}}
\newcolumntype{T}{>{\centering\arraybackslash}m{1.5cm}}
\newcolumntype{P}{>{\arraybackslash}m{7.0cm}}
\newcolumntype{H}{>{\arraybackslash}m{4cm}}
\newcolumntype{s}{>{\arraybackslash}m{0.8cm}}
\newcolumntype{C}[1]{>{\centering\arraybackslash}p{#1}}

\begin{document}
\title{Visualizing Self-Regulated Learner Profiles in Dashboards: Design Insights from Teachers}
\titlerunning{Visualizing Self-Regulated Learning Profiles}

\author{Paola Mejia-Domenzain  
\orcidID{0000-0003-1242-3134} \and
Eva Laini  \and
Seyed Parsa Neshaei \orcidID{0000-0002-4794-395X}  \and
Thiemo	Wambsganss  \and
Tanja Käser \orcidID{0000-0003-0672-0415} 
}
\authorrunning{P. Mejia-Domenzain et al.}
%
\institute{EPFL, Lausanne, Switzerland \\ 
\email{\{paola.mejia, seyed.neshaei, tanja.kaeser\}@epfl.ch} 
}

%
%
%
%
%
\maketitle              

\vspace{-5mm}
\begin{abstract}
Flipped Classrooms (FC) are a promising teaching strategy, where students engage with the learning material before attending face-to-face sessions. While pre-class activities are critical for course success, many students struggle to engage effectively in them due to inadequate of self-regulated learning (SRL) skills. Thus, tools enabling teachers to monitor students’ SRL and provide personalized guidance have the potential to improve learning outcomes. However, existing dashboards mostly focus on aggregated information, disregarding recent work leveraging machine learning (ML) approaches that have identified comprehensive, multi-dimensional SRL behaviors. Unfortunately, the complexity of such findings makes them difficult to communicate and act on. In this paper, we follow a teacher-centered approach to study how to make thorough findings accessible to teachers. We design and implement \texttt{FlippED}, a dashboard for monitoring students’ SRL behavior. We evaluate the usability and actionability of the tool in semi-structured interviews with ten university teachers. We find that communicating ML-based profiles spark a range of potential interventions for students and course modifications.

\end{abstract}
\vspace{-7mm}

\section{Introduction}



Over the past years, blended learning (BL), which combines in-person sessions with online learning, has gained increasing popularity. Students in \emph{Flipped Classroom (FC)} courses, a variation of BL, complete pre-class activities before participating in face-to-face sessions. While FCs have the potential to enhance student learning (e.g.,\cite{hardebolle2022gender}), they are not effective \textit{per se} \cite{saint2020, mejia-domenzain_identifying_2022}. Independently regulating their learning (e.g., managing their time) is a challenging task for many learners. While there are multiple existing solutions to monitor students' self-regulated learning (SRL) skills \cite{jivet2017awareness,ronald2022}, most of them overlook teachers' role in FCs and their ability to promote SRL skills and support students' learning experience \cite{perez2022designing}.

Recently, tools designed for teachers to support SRL (e.g., \cite{perez2022designing, Wiedbusch_Kite_Yang_Park_Chi_Taub_Azevedo_2021}) have been valued positively. For example, in MetaDash \cite{Wiedbusch_Kite_Yang_Park_Chi_Taub_Azevedo_2021}, teachers perceived the visualization of students' emotions as a valuable tool for lesson design. Furthermore, teachers appreciated the possibility to monitor student progress and engagement using a Moodle Plugin \cite{perez2022designing}. Most tools \cite{perez2022designing, Wiedbusch_Kite_Yang_Park_Chi_Taub_Azevedo_2021} visualize aggregated statistics of student behavior only, whereas recent work has demonstrated that students exhibit complex SRL patterns \cite{saint2020, mejia-domenzain_identifying_2022}. In particular, \cite{mejia-domenzain_identifying_2022} identified SRL profiles differing in levels of time management (regularity, effort, consistency, and proactivity) and metacognition (video control). These differences can get lost when aggregating them. Indeed,\cite{perez2022designing} found that providing aggregated information was not enough for supporting students. Therefore, machine learning (ML) approaches able to identify and represent comprehensive student behavior \cite{saint2020, mejia-domenzain_identifying_2022} build a promising basis for classroom orchestration.


Unfortunately, the findings of the aforementioned ML approaches are complex and therefore challenging to communicate and make accessible to teachers. The lack of trust in ML \cite{nazaretsky_instrument_2022, verbert2020learning} or unclear visualizations \cite{nazaretsky_empowering_2022, martinez2015latux} can hinder the adoption and use of ML-based teacher support tools.




In this paper, we therefore study how to best make comprehensive ML-based findings accessible to teachers in the context of SRL in FC settings. We identify students' multi-dimensional SRL profiles through clustering \cite{mejia-domenzain_identifying_2022} and design multiple visualizations based on information visualization findings \cite{gleicher2011visual,albers2014task}. We then implement our findings into \texttt{FlippED}, a teacher dashboard for monitoring students' SRL behavior. We investigate teachers' responses to the tool by conducting semi-structured interviews with ten university teachers. With our mixed-method approach, we aim to answer the following research questions: How do teachers interact with and respond to an ML-based tool for FC (\textbf{RQ1})? How actionable is the information provided in the dashboard (\textbf{RQ2})?

\vspace{-2mm}
\section{Teacher-Centered Design and Evaluation Framework}
\label{sec: methodology}
\vspace{-1mm}
To study how to communicate effectively ML-based findings to teachers, we followed a teacher-centered mixed-method approach. Requirements were
compiled both from the existing literature and from 10 user interviews. Then, we identified multi-dimensional SRL profiles of students in an introductory mathematics FC course~\cite{mejia-domenzain_identifying_2022}. Next, we designed visualizations of the identified patterns according to the nature of the data as well as prior work on visual designs \cite{gleicher2011visual,albers2014task, lee2016vlat}. Finally, we iterated the design with seven different teachers and evaluated the final dashboard with ten university teachers.

\vspace{-2mm}
\subsection{Multi-Dimensional SRL Profiles}
\vspace{-1mm}
We used the clustering framework suggested by \cite{mejia-domenzain_identifying_2022} to obtain multi-dimensional profiles of students' SRL behavior. The framework analyzes five dimensions of SRL identified as important for online learning in higher education \cite{broadbent2015self}: \textit{Effort} (intensity of student engagement), \textit{Consistency} (variation of effort over time), \textit{Regularity} (patterns of working days and hours), \textit{Proactivity} (anticipation or delay in course schedule), and \textit{Control} (control of cognitive load). The pipeline consists of two main steps. First, behavioral patterns for each dimension are obtained using Spectral Clustering. In a second step, the resulting labels per dimension are clustered using K-Modes.

We applied the pipeline to log data collected  from $292$ students ($29\%$ identifying as female and $71\%$ as male) of an undergraduate mathematics FC course~\cite{hardebolle2022gender}. After the first clustering step, we obtained different patterns for the five dimensions: for \textit{Effort} and \textit{Control}, difference in intensity (e.g., higher and lower intensity); for \textit{Consistency}, constant intensity and increase of intensity during exam preparation; for \textit{Proactivity}, up-to-date and delayed behavior; and for \textit{Regularity}, students with high or low regularity patterns. We then integrated the obtained patterns using the second clustering step and obtained five different profiles. For example, the best-performing profile had higher \textit{Control}, lower intensity \textit{Effort}, constant intensity (\textit{Consistency}), and up-to-date \textit{Proactivity}. The worst-performing profile had similar characteristics except the students were not up-to-date (\textit{Proactivity}).

A side effect of the richness of the analysis is that the profiles can be complex to understand. In the following, we explore how to best communicate the findings to empower teachers by providing easy-to-interpret and actionable information about their course, while maintaining a comprehensive analysis.

\vspace{-3mm}
\subsection{Teacher Dashboard}
\vspace{-1mm}
To address \textbf{RQ1} and \textbf{RQ2}, we designed a teacher dashboard displaying SRL profiles and behaviors. We evaluated the tool with ten teachers, analyzing their clickstream, think-aloud process, and semi-structured interview answers.

\begin{figure}[t]
    \centering
    \includegraphics[width=1\textwidth]{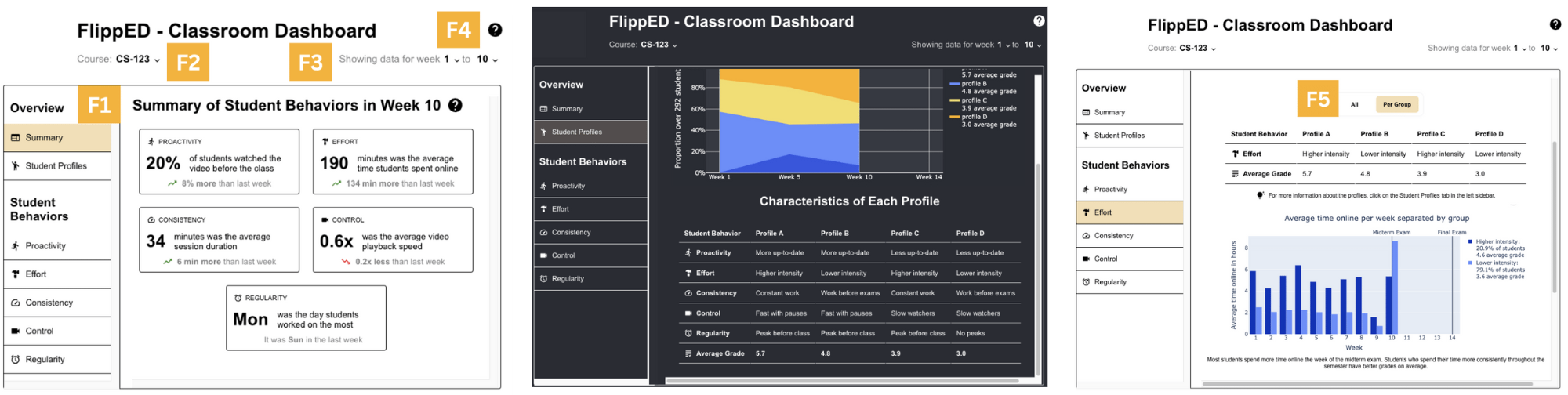}
    \caption{Example pages:  \textit{Summary},  \textit{Student Profiles}, and  \textit{Groups}.}
    \label{fig:overview}
     \vspace{-5mm}
\end{figure}

\vspace{1mm} \noindent \textbf{Design}. We designed and implemented \texttt{FlippED}, a teacher dashboard for SRL. We then iterated the design with seven pilot teachers. The final design includes a navigation menu (see F1 in Fig.~\ref{fig:overview}) with two parts: an overview displaying a \textit{Summary} and \textit{Student Profiles} and a detailed view illustrating students' behavioral patterns in the different dimensions. The user can select the course (F2) and the desired weeks (e.g., week $5-9$) (F3). In addition, there are help buttons (F4) throughout the dashboard providing further explanations. In the \textit{Summary} page (Fig.~\ref{fig:overview}-left), weekly statistics per dimension are displayed as well as the trend in comparison to the previous week. The description of the profiles and the associated grades are shown on the \textit{Profiles} page (Fig.~\ref{fig:overview}-center). Moreover, in the student behavior pages (Fig.~\ref{fig:overview}-right), users can choose (F5) between the aggregated view and a view per group.

\vspace{1mm} \noindent \textbf{Accessibility}. In order to make \texttt{FlippED} accessible to people with disabilities, we followed the guidelines from the \textit{Web Accessibility Initiative} (e.g., using a colorblind palette, providing detailed alt-texts and captions for all graphs, supporting dark contrast mode, and having a clear voice-navigable interface) and achieved the conformance level of AAA (highest level).


\vspace{1mm} \noindent \textbf{Participants}. We recruited ten university-level teachers ($50\%$ identified as female, 50\% as male) with experience in BL and FC through a faculty email.

\vspace{1mm} \noindent \textbf{Procedure}. We conducted semi-structured interviews asking participants to assume they were teaching a large FC course. The first task was to follow a think-aloud protocol and explore the dashboard. In the second task, participants were told that their students had just taken the midterm exam, and some students had surprisingly bad results. Thus, participants had to use the dashboard to give possible explanations. Lastly, we asked participants to assess whether and how they would use the dashboard in their classroom.
\section{Results}
\label{sec:results}

\subsection{Dashboard Usability (RQ1)}

To understand the usability of \texttt{FlippED}, in a first analysis, we investigated participants' exploration (task 1) clickstream together with their think-aloud comments. Participants started in the \textit{Summary} page and then either went into the \textit{Profiles} page or the \textit{Proactivity} page (the first menu item in \textit{Student Behaviors}, see Fig. \ref{fig:overview}). Then participants went back and forth to understand the layout and the profiles. Once they understood the structure of the dashboard, they followed mostly an ordered exploration strategy, accessing the pages following the sequential order from the menu.  Moreover, participants spent the longest time on the \textit{Profiles} page (on average $10$ minutes). Half of the participants were at first confused about the profiles but then said they would go back to the \textit{Profiles} page regularly and use them to advise students on the best learning strategies. 

In task 2, when asked to use the dashboard to identify possible causes for poor midterm performance, $80\%$ of the participants went straight to the \textit{Profiles} page and the remaining $20\%$ visited the \textit{Summary} page before the \textit{Profiles} page. Then, participants described the properties of the poor-performing profiles. 

\vspace{1mm} \noindent \textit{In summary, in the beginning, during the exploration task, participants viewed the student behavior pages (Proactivity, Effort, Consistency, Control, and Regularity) to understand each dimension and the observed behaviors. Then, during the second task, they were more drawn to the overview pages (Summary and Profiles) to get summarized information as a basis for suggesting interventions. }

\subsection{Actionability of Information (RQ2)}
In a second analysis, we examined the actionability of the provided information. Regarding the potential benefits of the dashboard, $80\%$ of the participants said they would show the students the dashboard in class. In particular, they mentioned adding the \textit{Proactivity} graphs to the course slides, showing the relationship between profile and grade to encourage proactive behaviors. As one participant illustrated, \textit{I would also use this information as feedback to students on their working habits. I would [show them this information] and say: 'Why don't you try to change this?'}.

In addition, $70\%$ of the interviewed teachers mentioned adapting the course in some way; some examples related to the \textit{Proactivity} dimension were asking students to come up with questions before the interactive sessions, adding activities such as continuous evaluation (quizzes), or proposing additional materials. Based on the \textit{Effort}, \textit{Consistency} and \textit{Regularity} pages, other possible actions teachers mentioned were sending motivational messages to all or some specific students, adapting the workload, sending automatic reminders to students who are not working regularly, and advising students in the lowest-grade profile to change their learning strategies. In addition, they proposed giving extra credit to promote watching videos and making the pre-class activities more entertaining.


\vspace{1mm} \noindent \textit{In summary, the visualizations helped teachers come up with a wide range of actionable items like adapting the course and communicating with the students.}


\vspace{-3mm}
\section{Implications and Conclusion}
\vspace{-2mm}
In this work, we studied the usability (\textbf{RQ1}) and actionability (\textbf{RQ2}) of ML-based teacher dashboards in the context of SRL in FC. We identified student behavioral profiles in FC using clickstream and communicated the findings in a teacher dashboard. Then, we evaluated its usability and actionability in semi-structured interviews with ten university teachers.

In contrast to existing teacher dashboards that focused on communicating aggregated statistics (e.g.,\cite{jivet2017awareness, perez2022designing, Wiedbusch_Kite_Yang_Park_Chi_Taub_Azevedo_2021}), we visualized and communicated intricate ML-based insights. Similar to \cite{nazaretsky_empowering_2022, wise2019teaching}, we found that most teachers judged the dashboard and visualizations as useful and actionable. In particular, participants appreciated the hierarchical design of the dashboard. They mentioned that the use of the overview pages (\textit{Summary} and \textit{Profiles}) displaying students' SRL profiles would be sufficient to design interventions on a weekly basis and used the detailed information on behavioral patterns mostly as a mean to understand the different SRL dimensions of the profiles. The hierarchical design of our dashboard solves two problems mentioned in previous work: not providing enough information for supporting student learning \cite{perez2022designing} and providing too much information \cite{verbert2020learning}. Furthermore, teachers perceived some dimensions (e.g., \textit{Proactivity}) as much more useful and actionable than others (e.g., \textit{Control}).




Our results are mostly consistent with prior work on teacher dashboards. In the following, we emphasize design guidelines for complex dashboards that go beyond those emphasized in earlier work:
\vspace{-1mm}
\begin{enumerate}[leftmargin=*]
    \item Structure the dashboard in a hierarchical way: include (1) overview and summary pages for daily use and (2) detailed information to gain a good understanding of the provided information. 
    \item Allow a flexible dashboard design that adjusts to the specific needs of the target teacher population. For example, omit dimensions that do not provide actionable items. 
\end{enumerate}
\vspace{-1mm}
\noindent Our work sheds insights into the interpretability, usability, and actionability of ML-based teacher dashboards. In the future, we plan to study the generalizability and validity of our findings in different contexts, regions, and cultures and to evaluate the usage of \texttt{FlippED} in diverse classrooms for longer periods.

\vspace{-3mm}
%
%
%
\bibliographystyle{splncs04}
\bibliography{bibliography}

\begin{thebibliography}{10}
\providecommand{\url}[1]{\texttt{#1}}
\providecommand{\urlprefix}{URL }
\providecommand{\doi}[1]{https://doi.org/#1}

\bibitem{albers2014task}
Albers, D., Correll, M., Gleicher, M.: Task-driven evaluation of aggregation in
  time series visualization. In: CHI (2014)

\bibitem{broadbent2015self}
Broadbent, J., Poon, W.L.: Self-regulated learning strategies \& academic
  achievement in online higher education learning environments: A systematic
  review. Internet High. Educ.  \textbf{27},  1--13 (2015)

\bibitem{gleicher2011visual}
Gleicher, M., Albers, D., Walker, R., Jusufi, I., Hansen, C.D., Roberts, J.C.:
  Visual comparison for information visualization. Inf Vis  \textbf{10}(4),
  289--309 (2011)

\bibitem{hardebolle2022gender}
Hardebolle, C., Verma, H., Tormey, R., Deparis, S.: Gender, prior knowledge,
  and the impact of a flipped linear algebra course for engineers over multiple
  years. J. Eng. Educ.  \textbf{111}(3),  554--574 (2022)

\bibitem{jivet2017awareness}
Jivet, I., Scheffel, M., Drachsler, H., Specht, M.: Awareness is not enough:
  Pitfalls of learning analytics dashboards in the educational practice. In:
  EC-TEL (2017)

\bibitem{lee2016vlat}
Lee, S., Kim, S.H., Kwon, B.C.: {VLAT}: Development of a visualization literacy
  assessment test. IEEE Trans Vis Comput Graph  \textbf{23}(1),  551--560
  (2016)

\bibitem{martinez2015latux}
Martinez-Maldonado, R., Pardo, A., Mirriahi, N., Yacef, K., Kay, J., Clayphan,
  A.: Latux: An iterative workflow for designing, validating, and deploying
  learning analytics visualizations. J. Learn. Anal.  \textbf{2}(3),  9--39
  (2015)

\bibitem{mejia-domenzain_identifying_2022}
Mejia-Domenzain, P., Marras, M., Giang, C., Käser, T.: Identifying and
  comparing multi-dimensional student profiles across flipped classrooms. In:
  AIED (2022)

\bibitem{nazaretsky_empowering_2022}
Nazaretsky, T., Bar, C., Walter, M., Alexandron, G.: Empowering teachers with
  {AI:} co-designing a learning analytics tool for personalized instruction in
  the science classroom. In: {LAK} (2022)

\bibitem{nazaretsky_instrument_2022}
Nazaretsky, T., Cukurova, M., Alexandron, G.: {An Instrument for Measuring
  Teachers’ Trust in {AI}-Based Educational Technology}. In: {LAK} (2022)

\bibitem{perez2022designing}
P{\'{e}}rez{-}Sanagust{\'{\i}}n, M., P{\'{e}}rez{-}{\'{A}}lvarez, R.,
  Maldonado{-}Mahauad, J., Villalobos, E., Sanza, C.: {Designing a Moodle
  Plugin for Promoting Learners' Self-regulated Learning in Blended Learning}.
  In: {EC-TEL} (2022)

\bibitem{saint2020}
Saint, J., Whitelock-Wainwright, A., Gašević, D., Pardo, A.: {Trace-SRL: A
  Framework for Analysis of Microlevel Processes of Self-Regulated Learning
  From Trace Data}. IEEE TLT  \textbf{13}(4),  861--877 (2020)

\bibitem{verbert2020learning}
Verbert, K., Ochoa, X., Croon, R.D., Dourado, R.A., Laet, T.D.: Learning
  analytics dashboards: the past, the present and the future. In: {LAK} (2020)

\bibitem{Wiedbusch_Kite_Yang_Park_Chi_Taub_Azevedo_2021}
Wiedbusch, M.D., Kite, V., Yang, X., Park, S., Chi, M., Taub, M., Azevedo, R.:
  {A Theoretical and Evidence-Based Conceptual Design of MetaDash: An
  Intelligent Teacher Dashboard to Support Teachers’ Decision Making and
  Students’ Self-Regulated Learning}. Frontiers in Education  \textbf{6},
  570229 (2021)

\bibitem{wise2019teaching}
Wise, A.F., Jung, Y.: Teaching with analytics: Towards a situated model of
  instructional decision-making. J. Learn. Anal.  \textbf{6}(2),  53--69 (2019)

\bibitem{ronald2022}
Álvarez, R.P., Jivet, I., Pérez-Sanagustín, M., Scheffel, M., Verbert, K.:
  Tools designed to support self-regulated learning in online learning
  environments: A systematic review. IEEE TLT  \textbf{15}(4),  508--522 (2022)

\end{thebibliography}

\end{document}